# Implementation of the 8-Nucleon Yakubovsky Formalism for Halo Nucleus $^8$He

E. Ahmadi Pouya*[1] & A. A. Rajabi[2]

*Physics Department, Shahrood University of Technology, P. O. Box 3619995161-316, Shahrood, Iran*

## Abstract

In order to study the bound-state structure of the Helium halo nuclei, the 8-nucleon Yakubovsky formalism has been implemented for $^8$He in a 5-body sub-cluster model, *i.e.* $\alpha + n + n + n + n$. In this case, the 8-nucleon Yakubovsky equations has been obtained in the form of two coupled equations, based on the two independent components. In addition, by removing the contribution interactions of the 8 and 7's bound nucleons in the formalism, the obtained equations explicitly reduce to the 6-nucleon Yakubovsky equations for $^6$He, in the case of effective 3-body model, *i.e.* $\alpha + n + n$. In view of the expectation for the dominant structure of $^8$He, namely an inert $\alpha$-core and four loosely-bound neutrons, Jacobi configurations of the two components in momentum space have been represented to provide technicalities which were considered useful for a numerical performance, such as bound-state calculations and momentum density distributions for halo-bound neutrons.

## Introduction

Studies of rich-neutron light nuclei present stringent tests for their calculations as they probe aspects of the interactions that are less prevalent in nuclei closer to stability. Among them, the physics of strong interactions gives rise to new structures in rich-neutron light nuclei. One of the best prominent cases are the helium halo nuclei, namely $^6$He and $^8$He with two and four loosely-bound neutrons, forming an extended halo around the $^4$He core. In this regard, the rich-neutron helium isotopes, $^6$He and $^8$He, are the topic of this study. Indeed, both nuclei are Borromean halos [1, 2], they have no bound excited states, and they exhibit the binding energy. Both helium halo nuclei are radioactive and decay quickly that complicates efforts to measure their properties. $^8$He has a time-life of only a tenth of a second and in this little time, can be considered as a stable nucleus. Practically, detection of the rare $^8$He is a major step for particle physicists. But, it is very important and interesting to understand how helium configures itself with strong interactions after being produced from a particle accelerator.

In recent decades, combination of nuclear and atomic physics methods has allowed new courses of precision measurements of the ground-state energies and charge radii of $^6$He [3, 4] and $^8$He [5, 6]. Dealing with the theoretical study of these nuclei is very challenging, because one needs to describe the small separation energy of the halo neutrons and the large radius of the whole system, in addition to the challenges of binding energy calculations. Because they are light-mass nuclei, one can use traditional methods to study them, such as Green's Function Monte Carlo (GFMC) method [7] and the No-Core Shell Model (NCSM) [8]. In addition, some of the standard methods in nuclear theory has been the progress of Effective Field Theory (EFT) and the Renormalization Group (RG) to nuclear forces. While such helium halo nuclei have been investigated using the NCSM [9], but still, no sense has been obtained for the halo structure properties in such theoretical studies.

In this paper, to approach a simple physical structure of the observed effects, namely $^8$He and $^6$He halo-bound nuclei, and to understand the physics that governs the neutron distributions in helium halos, it was felt that by using current/modern computers with high computational speed, beyond previous works [10-14], a further step to study the light halo nuclei, namely $^8$He as an 8-nucleon bound system problem, within the reliable Yakubovsky method [15] is now desirable. It is worth mentioning that our obtained results are more difficult to implement bound-state problem in few-body branch, due to its halo nature and high-dimensional calculations. However, we provide technicalities for numerical performances.

---
[1] E.Ahmady.ph@ut.ac.ir
[2] A.A.Rajabi@shahroodut.ac.ir

This paper is written as follows: In sect. I, the Yakubovsky formalism to the 8-nucleon bound system problem using the standard notation [16] is explicitly derived. In this case, the Lippmann-Schwinger scheme twice step-by-step has been applied on the integral form of Schrödinger equation, and then the identity of the nucleons are implemented. Finally, the 8-nucleon Yakubovsky formalism for effective $\alpha$-core structure in a 5-body sub-cluster model of $^8$He leads to a set of two coupled equations related to 2 different components. In sect. II, useful technicalities for numerical applications have been provided. To this aim, corresponding Jacobi configurations of each Yakubovsky component is schematically represented and then the integral form of coupled equations are represented by introducing basis states based on Jacobi momenta in partial-wave (PW) analysis. Finally, a summary in sect. III is given.

## I. The 8-nucleon Yakubovsky Formalism

In branch of few-body systems, each $N$-body bound system has $N(N-1)/2$ different two-body interactions. Therefore, in the 8-nucleon bound system there are 28 different two-nucleon interactions or 28 different cluster decompositions ($a_7$) having 7 clusters. They are indexed by the only pairing forces in cluster $a_7$ they contain, *e.g.* $a_7 = 12 \equiv 12 + 3 + 4 + 5 + 6 + 7 + 8$. In order to study the 8-nucleon bound system in the framework of the Yakubovsky method using the sub-cluster notation [16], the method is to first sum up the pair forces in each 7-body fragment ($a_7$), in a second step among all 6-body fragments ($a_6$), and then in a third step among all 5-body fragments ($a_5$). We cut off that formalism ending with 5-body sub-clusters in the spirit of the usually used approximating effective 5-body model that is $\alpha + n + n + n + n$. In the standard formalism, for an individual $\alpha$-particle, such two-body interactions, namely $\alpha - n$ and $\alpha$-clusters have not been used, because in this framework $\alpha$-core interacts as a four-nucleon sub-system in the 8-nucleon system. We start with the non-relativistic Schrödinger equation for the 8-nucleon bound system as

$$\left( H_0 + \sum_{a_7} V_{a_7} \right) \Psi = E\, \Psi, \tag{1.1}$$

where $H_0$ refers to the free Hamiltonian operator of the 8-nucleon system that will be introduced in the next section, and $\sum_{a_7} V_{a_7} \equiv V_{12} + \cdots + V_{78}$ is the summation of all paring interactions with 28 terms. According to the Lipmann-Schwinger scheme, Eq. (1.1) is rewritten into an integral equation

$$\Psi = G_0 \sum_{a_7} V_{a_7}\, \Psi, \tag{1.2}$$

where $G_0$ stand for the 8-nucleon free Green's function operator and in the case of bound states formalism we have $G_0 = [E - H_0]^{-1}$. Like previous works for five- and six-nucleon Yakubovsky formalism [12, 13], we start from the integral form of the Schrödinger equation, Eq. (1.2), applying Lipmann-Schwinger scheme twice step-by-step. We can define $\psi_{a_7} \equiv G_0 V_{a_7}\, \Psi$ by using Eqs. (1.2) and (1.7). Therefore, Faddeev like equation for 8-nucleon Yakubovsky amplitudes are given as

$$\psi_{a_7} \equiv G_0 t_{a_7} \sum_{b_7} \bar{\delta}_{a_7 b_7}\, \psi_{b_7}, \tag{1.3}$$

where $t_{a_7}$ is a pairing $t$-matrix operator that follows the Lippmann-Schwinger equation as $t_{a_7} = V_{a_7} + V_{a_7} G_0 t_{a_7}$ and $\bar{\delta}_{a_7 b_7} = 1 - \delta_{a_7 b_7}$. For the next step sub-cluster form the 6-body fragments amplitudes can be written

$$\psi_{a_7 a_6} = G_0 \sum_{b_7 \subset a_6} \mathcal{T}^{a_6}_{a_7 b_7} (\psi^{a_6})_{a_7} = G_0 \sum_{b_7 \subset a_6} \mathcal{T}^{a_6}_{a_7 b_7} \sum_{b_7 \subset b_6} \bar{\delta}_{a_6 b_6}\, \psi_{b_7 b_6}, \tag{1.4}$$

where $\mathcal{T}^{a_6}_{a_7 b_7}$ obeys as

$$\mathcal{T}^{a_6}_{a_7 b_7} = t_{a_7}\, \bar{\delta}_{a_7 b_7} + G_0 \sum_{c_7 \subset a_6} t_{a_7}\, \bar{\delta}_{a_7 c_7}\, \mathcal{T}^{a_6}_{c_7 b_7}, \tag{1.5}$$

next, more decompose the right side of Eq. (1.4) according to 5-body fragments are given as

$$\psi_{a_7,a_6}^{a_5} = G_0 \sum_{\substack{b_7 \subset a_6}} \mathcal{T}_{a_7 b_7}^{a_6} \sum_{\substack{b_7 \subset b_6 \\ b_6 \subset a_5}} \bar{\delta}_{a_6 b_6}\, \psi_{b_7 b_6}. \tag{1.6}$$

briefly, we work out that method ending with $(a_5)$ in the spirit of the applied though approximate effective $\alpha + n + n + n + n$ in a 5-body model. Further details could be found in Refs. [12, 13]. In the following, the single nucleons in sub-clusters, namely 7-, 6- and 5-body fragments, *i.e.* $(a_7)$, $(a_6)$ and $(a_5)$ respectively, will no longer be indexed.

Next, in order to get to 8-nucleon Yakubovsky coupled equations, the identity of the nucleons has been applied. In addition, the number of huge total WF Yakubovsky components has been obtained. It is well-known, the 8-nucleon total WF consists of the summation of all pairing Yakubovsky components with 28 terms which are as

$$\begin{aligned}\Psi = \sum_{a_7} \psi_{a_7} &\equiv \psi_{12} + \psi_{13} + \psi_{14} + \psi_{15} + \psi_{16} + \psi_{17} + \psi_{18} \\ &+ \psi_{23} + \psi_{24} + \psi_{25} + \psi_{26} + \psi_{27} + \psi_{28} \\ &+ \psi_{34} + \psi_{35} + \psi_{36} + \psi_{37} + \psi_{38} \\ &+ \psi_{45} + \psi_{46} + \psi_{47} + \psi_{48} \\ &+ \psi_{56} + \psi_{57} + \psi_{58} \\ &+ \psi_{67} + \psi_{68} \\ &+ \psi_{78},\end{aligned} \tag{1.7}$$

where the first pair $(\psi_{12})$, having 7-body fragments, consists of 21 components of 6-body fragments $(a_6)$ as

$$\begin{aligned}\psi_{a_7 \equiv 12} = \sum_{a_7 \equiv 12 \subset a_6} \psi_{a_7 \equiv 12, a_6} \\ \equiv \psi_{12,123} + \psi_{12,124} + \psi_{12,125} + \psi_{12,126} + \psi_{12,127} + \psi_{12,128} + \psi_{12,12+34} + \psi_{12,12+35} \\ + \psi_{12,12+36} + \psi_{12,12+37} + \psi_{12,12+38} + \psi_{12,12+45} + \psi_{12,12+46} + \psi_{12,12+47} + \psi_{12,12+48} \\ + \psi_{12,12+56} + \psi_{12,12+57} + \psi_{12,12+58} + \psi_{12,12+67} + \psi_{12,12+68} + \psi_{12,12+78},\end{aligned} \tag{1.8}$$

where $(a_6)$ refers to any 6-body fragments containing the pair restricted to $(a_7 \equiv 12)$ and the sum runs over pairs $a_7 \subset a_6$. It means that the sub-clusters $(a_6)$, when broken up, lead to the sub-clusters $(a_7)$. Next first term as $\psi_{12,123}$ is a sub-cluster of 15 components of 5-body fragments $(a_5)$

$$\begin{aligned}\psi_{a_7 \equiv 12, a_6 \equiv 123} = \sum_{a_6 \equiv 123 \subset a_5} \psi_{a_7 \equiv 12, a_6 \equiv 123}^{a_5} \\ \equiv \psi_{12,123}^{1234} + \psi_{12,123}^{1235} + \psi_{12,123}^{1236} + \psi_{12,123}^{1237} + \psi_{12,123}^{1238} + \psi_{12,123}^{123+45} + \psi_{12,123}^{123+46} + \psi_{12,123}^{123+47} \\ + \psi_{12,123}^{123+48} + \psi_{12,123}^{123+56} + \psi_{12,123}^{123+57} + \psi_{12,123}^{123+58} + \psi_{12,123}^{123+67} + \psi_{12,123}^{123+68} + \psi_{12,123}^{123+78},\end{aligned} \tag{1.9}$$

the summation of all 5-body fragments ends up to the 8-nucleon Yakubovsky components for total WF as

$$\Psi = \sum_{a_7} \sum_{a_6} \sum_{a_5} \psi_{a_7,a_6}^{a_5} = 8{,}820\, \psi_{a_7,a_6}^{a_5}, \tag{1.10}$$

We find that there are 8,820 components for the total WF of constituent 8 nucleons in a 5-body model. It is worthwhile to mention that to calculate the expectation value of the Hamiltonian operator, the above-mentioned total WF's have to be used. Clearly, the total WF is anti-symmetrized following the Pauli principle.

In the next step, the identity of the nucleons in sub-cluster components are implemented. We start from Eq. (1.4) choosing the case 7-body fragments $a_7 \equiv 12$ with 6-body fragments $a_6 \equiv 123$

$$\psi_{12,123} = G_0 \mathcal{T}^{123}_{12,12} \begin{pmatrix} \psi_{12,124} + \psi_{12,125} + \psi_{12,126} + \psi_{12,127} + \psi_{12,128} \\ +\psi_{12,12+34}+\psi_{12,12+35}+\psi_{12,12+36}+\psi_{12,12+37}+\psi_{12,12+38} \\ +\psi_{12,12+45}+\psi_{12,12+46}+\psi_{12,12+47}+\psi_{12,12+48} \\ +\psi_{12,12+56}+\psi_{12,12+57}+\psi_{12,12+58} \\ +\psi_{12,12+67}+\psi_{12,12+68} \\ +\psi_{12,12+78} \end{pmatrix}$$

$$+ G_0 \mathcal{T}^{123}_{12,23} \begin{pmatrix} \psi_{23,234} + \psi_{23,235} + \psi_{23,236} + \psi_{23,237} + \psi_{23,238} \\ +\psi_{23,23+14}+\psi_{23,23+15}+\psi_{23,23+16}+\psi_{23,23+17}+\psi_{23,23+18} \\ +\psi_{23,23+45}+\psi_{23,23+46}+\psi_{23,23+47}+\psi_{23,23+48} \\ +\psi_{23,23+56}+\psi_{23,23+57}+\psi_{23,23+58} \\ +\psi_{23,23+67}+\psi_{23,23+68} \\ +\psi_{23,23+78} \end{pmatrix}$$

$$+ G_0 \mathcal{T}^{123}_{12,31} \begin{pmatrix} \psi_{31,314} + \psi_{31,315} + \psi_{31,316} + \psi_{31,317} + \psi_{31,318} \\ +\psi_{31,31+24}+\psi_{31,31+25}+\psi_{31,31+26}+\psi_{31,31+27}+\psi_{31,31+28} \\ +\psi_{31,31+45}+\psi_{31,31+46}+\psi_{31,31+47}+\psi_{31,31+48} \\ +\psi_{31,31+56}+\psi_{31,31+57}+\psi_{31,31+58} \\ +\psi_{31,31+67}+\psi_{31,31+68} \\ +\psi_{31,31+78} \end{pmatrix},$$

(1.11)

it is easily seen that, by using adequate permutation operators, the huge equation above can be rewritten as

$$\psi_{12,123} = G_0 \left( \mathcal{T}^{123}_{12,12} + \mathcal{T}^{123}_{12,23} P_{12}P_{23} + \mathcal{T}^{123}_{12,31} P_{13}P_{23} \right)$$
$$\times \begin{pmatrix} \psi_{12,124} + \psi_{12,125} + \psi_{12,126} + \psi_{12,127} + \psi_{12,128} \\ +\psi_{12,12+34}+\psi_{12,12+35}+\psi_{12,12+36}+\psi_{12,12+37}+\psi_{12,12+38} \\ +\psi_{12,12+45}+\psi_{12,12+46}+\psi_{12,12+47}+\psi_{12,12+48} \\ +\psi_{12,12+56}+\psi_{12,12+57}+\psi_{12,12+58} \\ +\psi_{12,12+67}+\psi_{12,12+68} \\ +\psi_{12,12+78} \end{pmatrix},$$

(1.12)

by defining the bellow transition operator as
$$\mathcal{T}^{123} = \mathcal{T}^{123}_{12,12} + \mathcal{T}^{123}_{12,23} P_{12}P_{23} + \mathcal{T}^{123}_{12,31} P_{13}P_{23},$$

(1.13)

Eq. (1.12) is simplified to

$$\psi_{12,123} = G_0 \mathcal{T}^{123} \begin{pmatrix} \psi_{12,124} + \psi_{12,125} + \psi_{12,126} + \psi_{12,127} + \psi_{12,128} \\ +\psi_{12,12+34}+\psi_{12,12+35}+\psi_{12,12+36}+\psi_{12,12+37}+\psi_{12,12+38} \\ +\psi_{12,12+45}+\psi_{12,12+46}+\psi_{12,12+47}+\psi_{12,12+48} \\ +\psi_{12,12+56}+\psi_{12,12+57}+\psi_{12,12+58} \\ +\psi_{12,12+67}+\psi_{12,12+68} \\ +\psi_{12,12+78} \end{pmatrix}.$$

(1.14)

where $\mathcal{T}^{123}$ follows Lipmann-Schwinger equation as $\mathcal{T}^{123} = t_{12}P + t_{12}PG_0\mathcal{T}^{123}$ and $P = P_{12}P_{23} + P_{13}P_{23}$. Subsequently, starting again from Eq. (1.4) to select the case 7-body fragments $a_7 \equiv 12$ with 6-body fragments $a_6 \equiv 12 + 34$

$$\psi_{12,12+34} = G_0 \mathcal{T}_{12,12}^{12+34} \begin{pmatrix} \psi_{12,123} + \psi_{12,124} + \psi_{12,125} + \psi_{12,126} + \psi_{12,127} + \psi_{12,128} \\ +\psi_{12,12+35} + \psi_{12,12+36} + \psi_{12,12+37} + \psi_{12,12+38} \\ +\psi_{12,12+45} + \psi_{12,12+46} + \psi_{12,12+47} + \psi_{12,12+48} \\ +\psi_{12,12+56} + \psi_{12,12+57} + \psi_{12,12+58} \\ +\psi_{12,12+67} + \psi_{12,12+68} \\ +\psi_{12,12+78} \end{pmatrix}$$

$$+ G_0 \mathcal{T}_{12,34}^{12+34} \begin{pmatrix} \psi_{34,134} + \psi_{34,234} + \psi_{34,345} + \psi_{34,346} + \psi_{34,347} + \psi_{34,348} \\ +\psi_{34,34+15} + \psi_{34,34+16} + \psi_{34,34+17} + \psi_{34,34+18} \\ +\psi_{34,34+25} + \psi_{34,34+26} + \psi_{34,34+27} + \psi_{34,34+28} \\ +\psi_{34,34+56} + \psi_{34,34+57} + \psi_{34,34+58} \\ +\psi_{34,34+67} + \psi_{34,34+68} \\ +\psi_{34,34+78} \end{pmatrix}, \quad (1.15)$$

subsequently, by using permutation properties, Eq. (1.15) can be rewritten as

$$\psi_{12,12+34} = G_0 \left( \mathcal{T}_{12,12}^{12+34} + \mathcal{T}_{12,34}^{12+34} P_{13} P_{24} \right) \begin{pmatrix} \psi_{12,123} + \psi_{12,124} + \psi_{12,125} + \psi_{12,126} + \psi_{12,127} + \psi_{12,128} \\ +\psi_{12,12+35} + \psi_{12,12+36} + \psi_{12,12+37} + \psi_{12,12+38} \\ +\psi_{12,12+45} + \psi_{12,12+46} + \psi_{12,12+47} + \psi_{12,12+48} \\ +\psi_{12,12+56} + \psi_{12,12+57} + \psi_{12,12+58} \\ +\psi_{12,12+67} + \psi_{12,12+68} \\ +\psi_{12,12+78} \end{pmatrix}, \quad (1.16)$$

then defining

$$\mathcal{T}^{12+34} = \mathcal{T}_{12,12}^{12+34} + \mathcal{T}_{12,34}^{12+34} P_{13} P_{24}, \quad (1.17)$$

Eq. (1.16) is simplified to

$$\psi_{12,12+34} = G_0 \mathcal{T}^{12+34} \begin{pmatrix} \psi_{12,123} + \psi_{12,124} + \psi_{12,125} + \psi_{12,126} + \psi_{12,127} + \psi_{12,128} \\ +\psi_{12,12+35} + \psi_{12,12+36} + \psi_{12,12+37} + \psi_{12,12+38} \\ +\psi_{12,12+45} + \psi_{12,12+46} + \psi_{12,12+47} + \psi_{12,12+48} \\ +\psi_{12,12+56} + \psi_{12,12+57} + \psi_{12,12+58} \\ +\psi_{12,12+67} + \psi_{12,12+68} \\ +\psi_{12,12+78} \end{pmatrix}. \quad (1.18)$$

where $\mathcal{T}^{12+34}$ follows the equation $\mathcal{T}^{12+34} = t_{12}\tilde{P} + t_{12}\tilde{P}G_0\mathcal{T}^{12+34}$ and where $\tilde{P} = P_{13}P_{24}$.

The next step is decomposing $\psi_{12,123}$ according to Eq. (1.6). For 7-body fragments $a_7 \equiv 12$ with 6-body fragments $a_6 \equiv 123$ the all relevant 5-body fragments ($a_5$) are $1234, 1235, 1236, 1237, 1238, 123 + 45, 123 + 46, 123 + 47, 123 + 48, 123 + 56, 123 + 57, 123 + 58, 123 + 67, 123 + 68, 123 + 78$. All 5-body sub-clusters summed up

$$\psi_{12,123} = \psi_{12,123}^{1234} + \psi_{12,123}^{1235} + \psi_{12,123}^{1236} + \psi_{12,123}^{1237} + \psi_{12,123}^{1238} + \psi_{12,123}^{123+45} + \psi_{12,123}^{123+46} + \psi_{12,123}^{123+47} + \psi_{12,123}^{123+48} \quad (1.19)$$
$$+ \psi_{12,123}^{123+56} + \psi_{12,123}^{123+57} + \psi_{12,123}^{123+58} + \psi_{12,123}^{123+67} + \psi_{12,123}^{123+68} + \psi_{12,123}^{123+78}.$$

Considering Eq. (1.6), first part of the above turns

$$\psi_{12,123}^{1234} = G_0 \mathcal{T}_{12,12}^{123} (\psi_{12,124} + \psi_{12,12+34}) + G_0 \mathcal{T}_{12,23}^{123} (\psi_{23,234} + \psi_{23,23+14}) \quad (1.20)$$
$$+ G_0 \mathcal{T}_{12,31}^{123} (\psi_{31,134} + \psi_{31,31+24}),$$

since

$$\psi_{23,234} + \psi_{23,23+14} = P_{12}P_{23}(\psi_{12,124} + \psi_{12,12+34}), \quad (1.21)$$
$$\psi_{31,134} + \psi_{31,31+24} = P_{13}P_{23}(\psi_{12,124} + \psi_{12,12+34}), \quad (1.22)$$

according to Eq. (1.13), Eq. (1.20) leads to

$$\psi_{12,123}^{1234} = G_0 \mathcal{T}^{123}(\psi_{12,124} + \psi_{12,12+34}). \tag{1.23}$$

Similarly, next we decompose $\psi_{12,12+34}$ according to Eq. (1.6). For the 7-body fragments $a_7 \equiv 12$ with 6-body fragments $a_6 \equiv 12+34$ the possible 5-body fragments ($a_5$) are $1234, 125+34, 126+34, 127+34, 128+34, 12+345, 12+346, 12+347, 12+348, 12+34+56, 12+34+57, 12+34+58, 12+34+67, 12+34+68, 12+34+78$, which are now summarized to

$$\begin{aligned}\psi_{12,12+34} &= \psi_{12,12+34}^{1234} + \psi_{12,12+34}^{123+45} + \psi_{12,12+34}^{123+46} + \psi_{12,12+34}^{123+47} + \psi_{12,12+34}^{123+48} + \psi_{12,12+34}^{12+345} \\ &+ \psi_{12,12+34}^{12+346} + \psi_{12,12+34}^{12+347} + \psi_{12,12+34}^{12+348} + \psi_{12,12+34}^{12+34+56} + \psi_{12,12+34}^{12+34+57} + \psi_{12,12+34}^{12+34+58} \\ &+ \psi_{12,12+34}^{12+34+67} + \psi_{12,12+34}^{12+34+68} + \psi_{12,12+34}^{12+34+78}.\end{aligned} \tag{1.24}$$

Considering Eq. (1.6), the first part of the above equation is regarded in turn

$$\psi_{12,12+34}^{1234} = G_0 \mathcal{T}_{12,12}^{12+34}(\psi_{12,123} + \psi_{12,124}) + G_0 \mathcal{T}_{12,34}^{12+34}(\psi_{34,234} + \psi_{34,134}), \tag{1.25}$$

since we use $\psi_{34,234} + \psi_{34,134} = P_{13}P_{24}(\psi_{12,123} + \psi_{12,124})$. Eq. (1.25) is simplified according to Eq. (1.17), and rewritten as

$$\psi_{12,12+34}^{1234} = G_0 \mathcal{T}^{12+34}(\psi_{12,123} + \psi_{12,124}). \tag{1.26}$$

The two amplitudes, $\psi_{12,123}^{1234}$ and $\psi_{12,12+34}^{1234}$, expressed in Eq. (1.23) and Eq. (1.26) are coupled to each other as shown now. The expression Eq. (1.19) can easily be converted to $\psi_{12,124}$ and using in addition Eq. (1.24), with usage of Eq. (1.23), one obtains

$$\begin{aligned}\psi_{12,123}^{1234} = G_0 \mathcal{T}^{123} \Big( &(\psi_{12,124}^{1234} + \psi_{12,124}^{1245} + \psi_{12,124}^{1246} + \psi_{12,124}^{1247} + \psi_{12,124}^{1248} + \psi_{12,124}^{124+35} + \psi_{12,124}^{124+36} + \psi_{12,124}^{124+37} \\ &+ \psi_{12,124}^{124+38} + \psi_{12,124}^{124+36} + \psi_{12,124}^{124+57} + \psi_{12,124}^{124+58} + \psi_{12,124}^{124+67} + \psi_{12,124}^{124+68} + \psi_{12,124}^{124+78}) \\ &+ (\psi_{12,12+34}^{1234} + \psi_{12,12+34}^{123+45} + \psi_{12,12+34}^{123+46} + \psi_{12,12+34}^{123+47} + \psi_{12,12+34}^{123+48} + \psi_{12,12+34}^{12+345} \\ &+ \psi_{12,12+34}^{12+346} + \psi_{12,12+34}^{12+347} + \psi_{12,12+34}^{12+348} + \psi_{12,12+34}^{12+34+56} + \psi_{12,12+34}^{12+34+57} + \psi_{12,12+34}^{12+34+58} \\ &+ \psi_{12,12+34}^{12+34+67} + \psi_{12,12+34}^{12+34+68} + \psi_{12,12+34}^{12+34+78})\Big),\end{aligned} \tag{1.27}$$

correspondingly Eq. (1.26) yields

$$\begin{aligned}\psi_{12,12+34}^{1234} = G_0 \mathcal{T}^{12+34} \Big( &(\psi_{12,123}^{1234} + \psi_{12,123}^{1235} + \psi_{12,123}^{1236} + \psi_{12,123}^{1237} + \psi_{12,123}^{1238} + \psi_{12,123}^{123+45} \\ &+ \psi_{12,123}^{123+46} + \psi_{12,123}^{123+47} + \psi_{12,123}^{123+48} + \psi_{12,123}^{123+56} + \psi_{12,123}^{123+57} + \psi_{12,123}^{123+58} + \psi_{12,123}^{123+67} \\ &+ \psi_{12,123}^{123+68} + \psi_{12,123}^{123+78}) \\ &+ (\psi_{12,124}^{1234} + \psi_{12,124}^{1245} + \psi_{12,124}^{1246} + \psi_{12,124}^{1247} + \psi_{12,124}^{1248} + \psi_{12,124}^{124+35} + \psi_{12,124}^{124+36} + \psi_{12,124}^{124+37} \\ &+ \psi_{12,124}^{124+38} + \psi_{12,124}^{124+36} + \psi_{12,124}^{124+57} + \psi_{12,124}^{124+58} + \psi_{12,124}^{124+67} + \psi_{12,124}^{124+68} + \psi_{12,124}^{124+78})\Big),\end{aligned} \tag{1.28}$$

one separates now the components $\psi_{12,124}^{1234}$ and $\psi_{12,12+34}^{1234}$ in Eq. (1.27) from the rest

$$\psi_{12,123}^{1234} - G_0 \mathcal{T}^{123}\left(\psi_{12,124}^{1234} + \psi_{12,12+34}^{1234}\right)$$
$$= G_0 \mathcal{T}^{123}\big(\psi_{12,124}^{1245} + \psi_{12,124}^{1246} + \psi_{12,124}^{1247} + \psi_{12,124}^{1248} + \psi_{12,124}^{124+35} + \psi_{12,124}^{124+36} + \psi_{12,124}^{124+37}$$
$$+ \psi_{12,124}^{124+38} + \psi_{12,124}^{124+36} + \psi_{12,124}^{124+57} + \psi_{12,124}^{124+58} + \psi_{12,124}^{124+67} + \psi_{12,124}^{124+68} + \psi_{12,124}^{124+78}$$
$$+ \psi_{12,12+34}^{123+45} + \psi_{12,12+34}^{123+46} + \psi_{12,12+34}^{123+47} + \psi_{12,12+34}^{123+48} + \psi_{12,12+34}^{12+345} + \psi_{12,12+34}^{12+346} + \psi_{12,12+34}^{12+347}$$
$$+ \psi_{12,12+34}^{12+348} + \psi_{12,12+34}^{12+34+56} + \psi_{12,12+34}^{12+34+57} + \psi_{12,12+34}^{12+34+58} + \psi_{12,12+34}^{12+34+67} + \psi_{12,12+34}^{12+34+68}$$
$$+ \psi_{12,12+34}^{12+34+78}\big), \tag{1.29}$$

also, one separates now the components $\psi_{12,123}^{1234}$ and $\psi_{12,124}^{1234}$ in Eq. (1.28) from the rest

$$\psi_{12,12+34}^{1234} - G_0 \mathcal{T}^{12+34}\left(\psi_{12,123}^{1234} + \psi_{12,124}^{1234}\right)$$
$$= G_0 \mathcal{T}^{12+34}\big(\psi_{12,123}^{1235} + \psi_{12,123}^{1236} + \psi_{12,123}^{1237} + \psi_{12,123}^{1238} + \psi_{12,123}^{123+45} + \psi_{12,123}^{123+46} + \psi_{12,123}^{123+47}$$
$$+ \psi_{12,123}^{123+48} + \psi_{12,123}^{123+56} + \psi_{12,123}^{123+57} + \psi_{12,123}^{123+58} + \psi_{12,123}^{123+67} + \psi_{12,123}^{123+68} + \psi_{12,123}^{123+78}$$
$$+ \psi_{12,124}^{1245} + \psi_{12,124}^{1246} + \psi_{12,124}^{1247} + \psi_{12,124}^{1248} + \psi_{12,124}^{124+35} + \psi_{12,124}^{124+36} + \psi_{12,124}^{124+37} + \psi_{12,124}^{124+38}$$
$$+ \psi_{12,124}^{124+56} + \psi_{12,124}^{124+57} + \psi_{12,124}^{124+58} + \psi_{12,124}^{124+67} + \psi_{12,124}^{124+68} + \psi_{12,124}^{124+78}\big), \tag{1.30}$$

with $\psi_{12,124}^{1234} = -P_{34}\psi_{12,123}^{1234}$ Eqs. (1.29) and (1.30) into a matrix form

$$\begin{pmatrix}\psi_{12;123}^{1234}\\ \psi_{12;12+34}^{1234}\end{pmatrix} - G_0 \begin{pmatrix}\mathcal{T}^{123}(-P_{34}) & \mathcal{T}^{123}\\ \mathcal{T}^{12+34}(1-P_{34}) & 0\end{pmatrix}\begin{pmatrix}\psi_{12;123}^{1234}\\ \psi_{12;12+34}^{1234}\end{pmatrix}$$
$$= G_0 \begin{pmatrix}\mathcal{T}^{123}\begin{pmatrix}\psi_{12,124}^{1245} + \psi_{12,124}^{1246} + \psi_{12,124}^{1247} + \psi_{12,124}^{1248} + \psi_{12,124}^{124+35} + \psi_{12,124}^{124+36} + \psi_{12,124}^{124+37} + \psi_{12,124}^{124+38}\\ +\psi_{12,124}^{124+36} + \psi_{12,124}^{124+57} + \psi_{12,124}^{124+58} + \psi_{12,124}^{124+67} + \psi_{12,124}^{124+68} + \psi_{12,124}^{124+78}\\ +\psi_{12,12+34}^{123+45} + \psi_{12,12+34}^{123+46} + \psi_{12,12+34}^{123+47} + \psi_{12,12+34}^{123+48} + \psi_{12,12+34}^{12+345} + \psi_{12,12+34}^{12+346} + \psi_{12,12+34}^{12+347} + \psi_{12,12+34}^{12+348}\\ +\psi_{12,12+34}^{12+34+56} + \psi_{12,12+34}^{12+34+57} + \psi_{12,12+34}^{12+34+58} + \psi_{12,12+34}^{12+34+67} + \psi_{12,12+34}^{12+34+68} + \psi_{12,12+34}^{12+34+78}\end{pmatrix}\\ \mathcal{T}^{12+34}\begin{pmatrix}\psi_{12,123}^{1235} + \psi_{12,123}^{1236} + \psi_{12,123}^{1237} + \psi_{12,123}^{1238} + \psi_{12,123}^{123+45} + \psi_{12,123}^{123+46} + \psi_{12,123}^{123+47} + \psi_{12,123}^{123+48}\\ +\psi_{12,123}^{123+56} + \psi_{12,123}^{123+57} + \psi_{12,123}^{123+58} + \psi_{12,123}^{123+67} + \psi_{12,123}^{123+68} + \psi_{12,123}^{123+78}\\ +\psi_{12,124}^{1245} + \psi_{12,124}^{1246} + \psi_{12,124}^{1247} + \psi_{12,124}^{1248} + \psi_{12,124}^{124+35} + \psi_{12,124}^{124+36} + \psi_{12,124}^{124+37} + \psi_{12,124}^{124+38}\\ +\psi_{12,124}^{124+56} + \psi_{12,124}^{124+57} + \psi_{12,124}^{124+58} + \psi_{12,124}^{124+67} + \psi_{12,124}^{124+68} + \psi_{12,124}^{124+78}\end{pmatrix}\end{pmatrix}, \tag{1.31}$$

since

$$\psi_{12,124}^{1245} = -P_{34}\psi_{12,123}^{1235}, \quad \psi_{12,124}^{124+35} = -P_{34}\psi_{12,123}^{123+45}, \tag{1.32}$$

the right side of Eq. (1.31) can be factored and achieves the form as

$$\begin{pmatrix}\psi_{12;123}^{1234}\\ \psi_{12;12+34}^{1234}\end{pmatrix} - G_0 \begin{pmatrix}\mathcal{T}^{123}(-P_{34}) & \mathcal{T}^{123}\\ \mathcal{T}^{12+34}(1-P_{34}) & 0\end{pmatrix}\begin{pmatrix}\psi_{12;123}^{1234}\\ \psi_{12;12+34}^{1234}\end{pmatrix} = G_0 \begin{pmatrix}\mathcal{T}^{123}(-P_{34}) & \mathcal{T}^{123}\\ \mathcal{T}^{12+34}(1-P_{34}) & 0\end{pmatrix}$$
$$\times \begin{pmatrix}\begin{pmatrix}\psi_{12,123}^{1235} + \psi_{12,123}^{1236} + \psi_{12,123}^{1237} + \psi_{12,123}^{1238} + \psi_{12,123}^{123+45} + \psi_{12,123}^{123+46} + \psi_{12,123}^{123+47}\\ +\psi_{12,123}^{123+48} + \psi_{12,123}^{123+56} + \psi_{12,123}^{123+57} + \psi_{12,123}^{123+58} + \psi_{12,123}^{123+67} + \psi_{12,123}^{123+68} + \psi_{12,123}^{123+78}\end{pmatrix}\\ \begin{pmatrix}\psi_{12,12+34}^{123+45} + \psi_{12,12+34}^{123+46} + \psi_{12,12+34}^{123+47} + \psi_{12,12+34}^{123+48} + \psi_{12,12+34}^{12+345} + \psi_{12,12+34}^{12+346} + \psi_{12,12+34}^{12+347}\\ +\psi_{12,12+34}^{12+348} + \psi_{12,12+34}^{12+34+56} + \psi_{12,12+34}^{12+34+57} + \psi_{12,12+34}^{12+34+58} + \psi_{12,12+34}^{12+34+67} + \psi_{12,12+34}^{12+34+68} + \psi_{12,12+34}^{12+34+78}\end{pmatrix}\end{pmatrix}, \tag{1.33}$$

the right side can be reduced applying permutation operators and obtains the final form of the 8-nucleon equations as

$$\begin{pmatrix} \psi_{12;123}^{1234} \\ \psi_{12;12+34}^{1234} \end{pmatrix} = G_0 \begin{pmatrix} \mathcal{T}^{123}(-P_{34}) & \mathcal{T}^{123} \\ \mathcal{T}^{12+34}(1-P_{34}) & 0 \end{pmatrix} \quad (1.34)$$
$$\times \left[ \begin{pmatrix} (1-P_{45}-P_{46}-P_{47}-P_{48})\psi_{12;123}^{1234} + (1-P_{56}-P_{57}-P_{58}-P_{46}-P_{47}-P_{48}+P_{46}P_{57}+P_{46}P_{58}+P_{47}P_{57})\psi_{12;123}^{123+45} \\ \psi_{12;12+34}^{1234} + (1-P_{56}-P_{57}-P_{58})(\psi_{12;12+34}^{125+34} + \psi_{12;12+34}^{12+345}) + (1-P_{67}-P_{68}-P_{57}-P_{58}+P_{57}P_{68})\psi_{12,12+34}^{12+34+56} \end{pmatrix} \right],$$

we end the formalism with some Yakubovsky independent components as $\psi_{12;123}^{1234}$, $\psi_{12;12+34}^{1234}$, $\psi_{12;123}^{123+45}$, $(\psi_{12,12+34}^{125+34} + \psi_{12,12+34}^{12+345})$ and $\psi_{12,12+34}^{12+34+56}$ coupled in the equations. According to the 5-body sub-cluster distributions of above independent components, it is well-known that two first components are related to $4N + N + N + N + N$ that refer to approximately effective 5-body model ($\alpha + n + n + n + n$). Respectively, third component refers to $3N + 2N + N + N + N$, which is not present in the effective 5-body $\alpha$-core model. Fourth component is a linear combinations that refer again to the third component. This is also beyond the effective 5-body $\alpha$-core model. The last one is related to $2N + 2N + 2N + 2N$ sub-cluster which is not contained in the effective 5-body $\alpha$-core model, either. Regarding the sub-cluster underlying the Yakubovsky components, only $\psi_{12;123}^{1234}$ and $\psi_{12;12+34}^{1234}$, are related to approximately effective configuration of $\alpha + n + n + n + n$, where the $\alpha$-core and four loosely bound neutrons approximation are valid. Now, here we convinced enough why we choose two specific components, and those components are related to the effective $\alpha$-core structure. By considering the distinct Jacobi configurations related to remaining Yakubovsky components, Fig.1, we can choose the first two components, $\psi_{12;123}^{1234}$ and $\psi_{12;12+34}^{1234}$, and obviously other components will not be taken into account.

Hint: after removing the contribution interactions of the seventh and eighth nucleons in Eq. (1.34), *i.e.* removing their permutation operators, the 8-nucleon obtained equations leads to the 6-nucleon ones for halo nucleus $^6$He (See Eq. (67) in Ref. [12]). Such reduction, confirms that our implementing the Yakubovsky formalism for the 8-nucleon bound system for $^8$He, is appropriate approximation to describe the effective $\alpha$-core structure for Helium halo-bound nuclei, similar to the approximation of the six-nucleon for $^6$He in the effective 3-body model [12, 13].

In the above explanations we clearly have discussed why only the two specific components, $\psi_{12;123}^{1234}$ and $\psi_{12;12+34}^{1234}$, are relevant to the effective $\alpha$-core structure of $^8$He, and the other Yakubovsky components in Eq. (1.34), will not be taken into account. Therefore, Eq. (1.34) leads to

$$\begin{pmatrix} \psi_{12;123}^{1234} \\ \psi_{12;12+34}^{1234} \end{pmatrix} = G_0 \begin{pmatrix} \mathcal{T}^{123}(-P_{34}) & \mathcal{T}^{123} \\ \mathcal{T}^{12+34}(1-P_{34}) & 0 \end{pmatrix} \left[ \begin{pmatrix} (1-P_{45}-P_{46}-P_{47}-P_{48})\psi_{12;123}^{1234} \\ \psi_{12;12+34}^{1234} \end{pmatrix} \right]. \quad (1.35)$$

It is worth mentioning that switching off some Yakubovsky independent components changes the Hamiltonian of the comprehensive 8-nucleon system, so the Hamiltonian does not contain all interactions, but we show that in the remaining components the effective interactions of $\alpha$-core as the stable 4-nucleon sub-system is governed in the remaining components (Refer to Fig.1). It is obvious that after removing the contribution interactions of the 5 to 8's nucleons in the 8-nucleon Yakubovsky equations, Eq. (1.34) straightforwardly reduces to the four-nucleon Yakubovsky equations for $^4$He ( See Ref. [10]) as

$$\begin{pmatrix} \psi_{12;123}^{1234} \\ \psi_{12;12+34}^{1234} \end{pmatrix} = G_0 \begin{pmatrix} \mathcal{T}^{123}(-P_{34}) & \mathcal{T}^{123} \\ \mathcal{T}^{12+34}(1-P_{34}) & 0 \end{pmatrix} \begin{pmatrix} \psi_{12;123}^{1234} \\ \psi_{12;12+34}^{1234} \end{pmatrix}, \quad (1.36)$$

in order to reduce the elaboration of the 8-nucleon coupled equations, Eq. (1.35), we can present the Yakubovsky equations for the 8-nucleon problem as a function of two-nucleon $t$-matrices and avoid using $\mathcal{T}^{123}$ and $\mathcal{T}^{12+34}$. Consequently, the formalism can be simplified and the numerical solution of integral would be much faster in comparison with the case where was need to solve first sub-cluster Faddeev-like equation to obtain $\mathcal{T}^{123}$ and $\mathcal{T}^{12+34}$ by using Padé′ approximation [10]. In addition to Eq. (1.36), the 4-nucleon Yakubovsky coupled equations has been obtained in [10] as

$$\begin{pmatrix} \psi_1 \\ \psi_2 \end{pmatrix} = \begin{pmatrix} G_0 t_{12} P(1 - P_{34}) & G_0 t_{12} P \\ G_0 t_{12} \tilde{P}(1 - P_{34}) & G_0 t_{12} \tilde{P} \end{pmatrix} \begin{pmatrix} \psi_1 \\ \psi_2 \end{pmatrix}. \tag{1.37}$$

also in Eq. (1.36) the 4-nucleon transition operators obey as

$$TP = t_{12}P + PG_0 t_{12}TP \equiv \mathcal{T}^{123}, \tag{1.38}$$

$$\tilde{T}\tilde{P} = t_{12}\tilde{P} + \tilde{P}G_0 t_{12}\tilde{T}\tilde{P} \equiv \mathcal{T}^{12+34}, \tag{1.39}$$

According to [10] the 4-nucleon matrix form transition in Eq. (1.36) and (1.37) are similar in action, and according to Eqs. (1.35) and (1.36), the 8-nucleon and the 4-nucleon matrix form transitions are equal in action. At this point, the matrices of Eq. (1.37) can be identified and Eq. (1.35) can be written in terms of two-nucleon $t-$matrices operator since they are given as

$$\begin{pmatrix} \psi^{1234}_{12;123} \\ \psi^{1234}_{12;12+34} \end{pmatrix} = \begin{pmatrix} G_0 t_{12} P(1 - P_{34}) & G_0 t_{12} P \\ G_0 t_{12} \tilde{P}(1 - P_{34}) & G_0 t_{12} \tilde{P} \end{pmatrix} \left[ \begin{pmatrix} (1 - P_{45} - P_{46} - P_{47} - P_{48})\psi^{1234}_{12,123} \\ \psi^{1234}_{12;12+34} \end{pmatrix} \right], \tag{1.40}$$

finally, the linear form of the obtained 8-nucleon Yakubovsky equations in the case of effective $\alpha$-core structure can be written as

$$\psi^{1234}_{12;123} = G_0 t_{12} P\big[(1 - P_{34})(1 - P_{45} - P_{46} - P_{47} - P_{48})\psi^{1234}_{12;123} + \psi^{1234}_{12;12+34}\big], \tag{1.41}$$

$$\psi^{1234}_{12;12+34} = G_0 t_{12} \tilde{P}\big[(1 - P_{34})(1 - P_{45} - P_{46} - P_{47} - P_{48})\psi^{1234}_{12;123} + \psi^{1234}_{12;12+34}\big]. \tag{1.42}$$

In the next step we will describe characteristic numerical techniques such as introducing Jacobi configuration of each independent component in momentum space representation to provide corresponding basis states and to understand the halo structure model of the 8-nucleon system with respect to the regarded relevant remaining components; integral representation of the coupled equations; eigenvalue equation form and iteration method for solution of it, and discussions about the halo structure configurations that describe the 8-nucleon halo-bound system in the case of effective $\alpha$-core structure.

## II.  Technicalities for Numerical Applications

In this section, first, Jacobi configurations in momentum space are represented to define corresponding momentum basis states for evaluations of the integral kernels. Next, details for a typical algorithm are described to solve coupled integral equations based on PW representation. In order to describe the corresponding momentum basis states, for two independent components of the dominant $\alpha$-core structure of $^8$He, namely $\psi^{1234}_{12;123}$ and $\psi^{1234}_{12;12+34}$ the standard Jacobi momenta are described in Fig. 1, respectively.

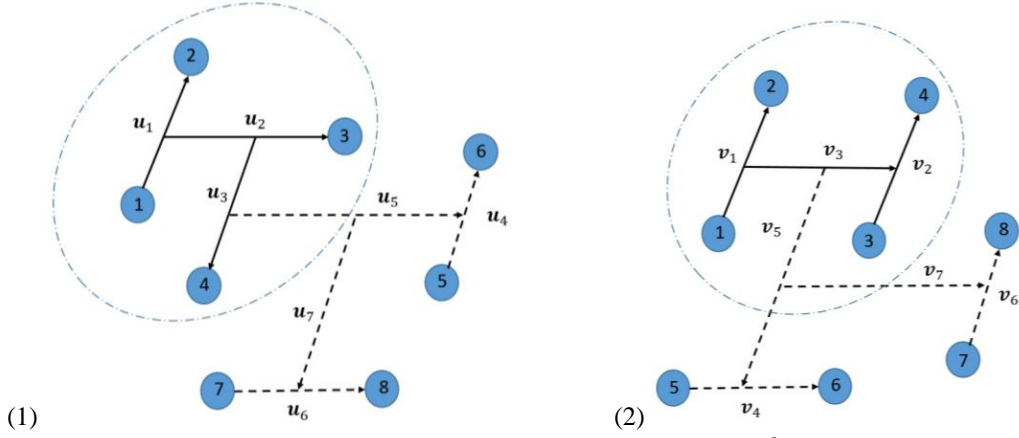

Fig. 1: Schematic representation of the two independent components of 8-nucleon bound system for $^8$He (four loosely bound neutrons with an inert $\alpha$-core) in Jacobi configurations. In the bag the $\alpha$-core tightly interacts in a 4-nucleon subsystem.

For the first component $\psi^{1234}_{12;123}$ in terms of the first configuration in Fig. 1 sets corresponding to $\boldsymbol{u}$-set chains

$$\begin{pmatrix}\boldsymbol{u}_1\\ \boldsymbol{u}_2\\ \boldsymbol{u}_3\\ \boldsymbol{u}_4\\ \boldsymbol{u}_5\\ \boldsymbol{u}_6\\ \boldsymbol{u}_7\\ U\end{pmatrix}=\begin{pmatrix}1/2 & -1/2 & 0 & 0 & 0 & 0 & 0 & 0\\ -1/3 & -1/3 & 2/3 & 0 & 0 & 0 & 0 & 0\\ -1/4 & -1/4 & -1/4 & 3/4 & 0 & 0 & 0 & 0\\ 0 & 0 & 0 & 0 & 1/2 & -1/2 & 0 & 0\\ -1/3 & -1/3 & -1/3 & -1/3 & 2/3 & 2/3 & 0 & 0\\ 0 & 0 & 0 & 0 & 0 & 0 & 1/2 & -1/2\\ -1/4 & -1/4 & -1/4 & -1/4 & -1/4 & -1/4 & 3/4 & 3/4\\ 1 & 1 & 1 & 1 & 1 & 1 & 1 & 1\end{pmatrix}\begin{pmatrix}\boldsymbol{p}_1\\ \boldsymbol{p}_2\\ \boldsymbol{p}_3\\ \boldsymbol{p}_4\\ \boldsymbol{p}_5\\ \boldsymbol{p}_6\\ \boldsymbol{p}_7\\ \boldsymbol{p}_8\end{pmatrix}, \quad (2.1)$$

in the non-relativistic case, we can define the kinetic energy operator by two equivalent forms. The inverse form of the above-mentioned transfer matrix is used to represent the kinetic energy in terms of $\boldsymbol{u}$-set Jacobi momenta.

$$H_0^u=\sum_{i=1}^{8}\frac{p_i^2}{2m}\equiv\frac{u_1^2}{m}+\frac{3}{4}\frac{u_2^2}{m}+\frac{2}{3}\frac{u_3^2}{m}+\frac{u_4^2}{m}+\frac{3}{8}\frac{u_5^2}{m}+\frac{u_6^2}{m}+\frac{1}{3}\frac{u_7^2}{m}, \quad (2.2)$$

where $\boldsymbol{p}_i$ is an individual particle momentum in the center of mass situation (under the condition $\sum_i \boldsymbol{p}_i=0$) that is described by relative Jacobi momenta $\boldsymbol{u}_i$; ($i=1,2,3,4,5,6,7,8$). Similarly, to $\psi^{1234}_{12;12+34}$ in terms of the second configuration in Fig. 1 belongs

$$\begin{pmatrix}\boldsymbol{v}_1\\ \boldsymbol{v}_2\\ \boldsymbol{v}_3\\ \boldsymbol{v}_4\\ \boldsymbol{v}_5\\ \boldsymbol{v}_6\\ \boldsymbol{v}_7\\ V\end{pmatrix}=\begin{pmatrix}1/2 & -1/2 & 0 & 0 & 0 & 0 & 0 & 0\\ 0 & 0 & 1/2 & -1/2 & 0 & 0 & 0 & 0\\ -1/2 & -1/2 & 1/2 & 1/2 & 0 & 0 & 0 & 0\\ 0 & 0 & 0 & 0 & 1/2 & -1/2 & 0 & 0\\ -1/3 & -1/3 & -1/3 & -1/3 & 2/3 & 2/3 & 0 & 0\\ 0 & 0 & 0 & 0 & 0 & 0 & 1/2 & -1/2\\ -1/4 & -1/4 & -1/4 & -1/4 & -1/4 & -1/4 & 3/4 & 3/4\\ 1 & 1 & 1 & 1 & 1 & 1 & 1 & 1\end{pmatrix}\begin{pmatrix}\boldsymbol{p}_1\\ \boldsymbol{p}_2\\ \boldsymbol{p}_3\\ \boldsymbol{p}_4\\ \boldsymbol{p}_5\\ \boldsymbol{p}_6\\ \boldsymbol{p}_7\\ \boldsymbol{p}_8\end{pmatrix}, \quad (2.3)$$

according to some appearing Jacobi vectors in the two different configurations in Fig.1, correspondingly, some equal Jacobi momenta have been represented by different notations, *i.e.* $\boldsymbol{v}_7=\boldsymbol{u}_7$, because of the standard representation of the above transfer matrices, for providing inverse transformations. The kinetic energy form in terms of $\boldsymbol{v}$-set Jacobi momenta, are given as

$$H_0^v = \sum_{i=1}^{8} \frac{p_i^2}{2m} \equiv \frac{v_1^2}{m} + \frac{v_2^2}{m} + \frac{1}{2}\frac{v_3^2}{m} + \frac{v_4^2}{m} + \frac{3}{8}\frac{v_4^2}{m} + \frac{v_4^2}{m} + \frac{1}{3}\frac{v_4^2}{m}. \tag{2.4}$$

Then, we introduce the two basis-states corresponding two independent components in PW representation. Obviously, these basis are separated into the 3 spaces that are 1-momentum/ 2-angular momentum and spin/ 3-isospin. The relevant basis-states in PW representation for $\psi_{12;123}^{1234}$ and $\psi_{12;12+34}^{1234}$ respectively are given as

$$|u; \gamma_u\rangle \equiv |u_1 u_2 u_3 u_4 u_5 u_6 u_7\rangle$$
$$\otimes \left|(l_1 s_{12})j_1 \left(l_2 \frac{1}{2}\right) j_2 (j_1 j_2) I_3 \left(l_3 \frac{1}{2}\right) j_4 (I_3 j_4) I_4 (l_5 s_{56}) j_5 (l_4 j_5) I_5 (I_4 I_5) I_6 (l_6 s_{12}) j_6 (l_7 j_6) I_7 (I_6 I_7) J, M_j\right\rangle$$
$$\otimes \left|(t_{12} \frac{1}{2}) t_3 \left(t_3 \frac{1}{2}\right) t_4 (t_4 t_{56}) t_6 (t_6 t_{78}) T, M_t\right\rangle, \tag{2.5}$$

$$|v; \gamma_v\rangle \equiv |v_1 v_2 v_3 v_4 v_5 v_6 v_7\rangle$$
$$\otimes |(l_1 s_{12}) j_1 (l_2 s_{34}) j_2 (j_1 j_2) S(L\ S) I(l_4 s_{56}) j_5 (l_5 j_5) I_5 (I I_5) I_6 (l_4 s_{78}) j_6 (l_7 j_6) I_7 (I_6 I_7) J, M_j\rangle$$
$$\otimes |(t_{12} t_{34}) t_{1\_4} (t_{1\_4} t_{56}) t_6 (t_6 t_{78}) T, M_t\rangle, \tag{2.6}$$

here the first part in the above-introduced basis states are magnitude of momentum vectors in PW representation. In the second part the orbital angular momentum $l_i$ go with the $\boldsymbol{u}_i\backslash \boldsymbol{v}_i$, $s_{ij}$ which are two-body spins for nucleons $ij$, $j_i$ are total 1- and 2-body angular momenta coupled out of orbital and spin angular momenta, $I_i$ are total $i$-body angular momenta except $I_5$ and $I_7$. These are total angular momenta of 5, 6 and 7, 8 respectively. Evidently, the third part refers to isospin. In this step, in order to implement the numerical techniques, the two coupled equations, Eq. (1.41) and Eq. (1.42), by inserting the completeness relations between the permutation operators, can be projected in momentum space as

$$\langle u; \gamma_u | \psi_{12;123}^{1234}\rangle = \sum_{\gamma_{u'}} \int u'^2 du' \sum_{\gamma_{u''}} \int u''^2 du''$$
$$\langle u; \gamma_u | G_0 t_{12} P | u'; \gamma_{u'}\rangle\langle u'; \gamma_{u'} | (1-P_{34})(1-P_{45}-P_{46}-P_{47}-P_{48}) | u''; \gamma_{u''}\rangle\langle u''; \gamma_{u''} | \psi_{12;123}^{1234}\rangle$$
$$+ \sum_{\gamma_{u'}} \int u'^2 du' \sum_{\gamma_{v'}} \int v'^2 dv' \langle u; \gamma_u | G_0 t_{12} P | u'; \gamma_{u'}\rangle\langle u'; \gamma_{u'} | v'; \gamma_{v'}\rangle\langle v'; \gamma_{v'} | \psi_{12;12+34}^{1234}\rangle, \tag{2.7}$$

$$\langle v; \gamma_v | \psi_{12;12+34}^{1234}\rangle = \sum_{\gamma_{v'}} \int v'^2 dv' \sum_{\gamma_{u'}} \int u'^2 du'$$
$$\langle v; \gamma_v | G_0 t_{12} \tilde{P} | v'; \gamma_{v'}\rangle\langle v'; \gamma_{v'} | (1-P_{34})(1-P_{45}-P_{46}-P_{47}-P_{48}) | u'; \gamma_{u'}\rangle\langle u'; \gamma_{u'} | \psi_{12;123}^{1234}\rangle$$
$$+ \sum_{\gamma_{v'}} \int v'^2 dv' \langle v; \gamma_v | G_0 t_{12} \tilde{P} | v'; \gamma_{v'}\rangle\langle v'; \gamma_{v'} | \psi_{12;12+34}^{1234}\rangle, \tag{2.8}$$

using techniques like ones presented in Refs. [10-14], it is straightforward to evaluate in PW analysis like the above-mentioned kernels and shifted momenta. It is well-known that the variables for 7 amplitudes on the right side are in general linear combinations of intermediate integration variables which include angles besides momentum magnitudes. After evaluating each term in the above-mentioned coupled integral equations in the standard PW analysis, the obtained equations are the starting point for numerical calculations in the eigenvalue equation form. We can replace the continuous variables in the numerical treatment by a dependence on certain distinct values by using the Gauss-Legendre discretization. The eigenvalue equation can be solved by the iteration method. It can apply Lanczos-like scheme that is efficient for nuclear few-body problems [10]. Also the evaluated coupled integral equations involve a very large number of interpolations, it can be used the cubic Hermit splines of Ref. [17] for its accuracy and high computational speed. The eigenvalue equation form can be given as

$$\eta(E)\,\psi(\psi_{12;123}^{1234},\psi_{12;12+34}^{1234}) = k(E)\,\psi(\psi_{12;123}^{1234},\psi_{12;12+34}^{1234}), \qquad (2.9)$$

where $E$ is the energy eigenvalue at which auxiliary Yakubovsky kernel eigenvalue is $\eta(E) = 1$. Because the energy is varied such that one reaches eigenvalue to be one. The Yakubovsky kernel of the equations $k(E)$ is just depend to energy $E$, and $\eta(E)$ is eigenvalue with $\psi(\psi_{12;123}^{1234},\psi_{12;12+34}^{1234})$ as the corresponding eigenvector. In order to solve the eigenvalue equation, Eq. (2.9), it can be used the Gaussian quadrature grid points. The coupled equations represent a set of homogenous integral equations, which turn into a high-dimension matrix eigenvalue equation after discretization. Starting from an arbitrary initial $\psi \equiv \psi_0$ one generates by consecutive applications of $k(E)$ a sequence of amplitudes $\psi_n$, which after orthogonalization form a basis into which $\psi$ is expanded. Then the energy is varied such that one reaches $\eta(E) = 1$. In order to investigate the loosely interactions of the halo bound neutrons in $^8$He, calculations of the momentum probability density as $n(u_i)[fm]$ in terms of $u_i[fm^{-1}]$, correspondingly ($v_i$) can be done. Why we just select 4 and 6 Jacobi momenta to calculate the momentum probability density? Because they are the only ones that make up contributions of the halo neutrons (refer to Eqs. (2.1) and (2.3)). By calculating momentum probability densities, we can represent the situation of the halo bound neutrons in momentum space. It is expected that the representation of momentum probability densities approximately are wide and short in comparison with tightly bound neutrons, namely neutrons inside the $\alpha$-core subsystem. Such those momentum distributions confirms that the 8-nucleon in the case of effective $\alpha$-core model is a halo structure system, exactly like halo nucleus $^8$He.

## III. Summary

The applications of Faddeev and Yakubovsky equations are traditionally for 3- and 4-nucleon bound and scattering states. Today, after the experiences with 4-, 5- and 6-nucleon bound-state problems within the Yakubovsky approach [13, 14] that the technical expertise has developed in the recent years and the very strong increase of computer power just recently achieved, it is possible to approach the 8-nucleon bound-state problem in that formalism and then numerical solution. For a full solution of the 8-nucleon Yakubovsky equations for $^8$He, modern super-computers with parallel algorithms is required. But in order to approach the calculations of the 8-nucleon bound system, for the first time, we have implemented the 8-nucleon Yakubovsky equations for the halo nucleus $^8$He in the case of simple effective $\alpha$-core structure. To this end, we stop the sequential sub-clustering with 5 fragments, namely $\alpha + n + n + n + n$, though the additional step with 4 fragments could be easily performed, but is not applicable for a physical system. In this method, we restricted the formulation to two-nucleon forces only because loosely bound neutrons with the center of an inert $\alpha$-core tend to have much stronger pairing interactions for large radii neutrons [5]. Therefore, as a simplification in the formalism, and above fact in such a model, three-nucleon forces was ignored. As a result, the implementation of the 8-nucleon bound system in the halo structure model within the Yakubovsky approach leads to two coupled equations in terms of two independent Yakubovsky components. Subsequently, in order to approach the halo structures of $^8$He, in addition to the analytical implementations, we have introduced the Jacobi configurations of those components in momentum space representation. Eventually, we provide technicalities which were considered useful for a numerical performance, such as bound-state calculations and proving the halo structure of bound neutrons that are outside of the $\alpha$-core.